# A Punch line: altermagnetism

*The first draft of the editorial introduction to the Perspective article by L. Smejkal et al.*

*I.I. Mazin*

An Irish Franciscan monk John Punch, in 1639, observed that *Non sunt multiplicanda entia sine necessitate,* entities should not be multiplied without necessity[1]. PRX is committed to follow this rule and weed out from its publications unnecessary terminology, invented solely as an advertisement tool. Yet, we had solicited and are publishing the following Perspective that seemingly does exactly that: introduces a new name in the decades-old field of magnetic materials.

Interesting magnetic patterns have been subject of intense research in the last decades, not in the least because of growing interest in spintronics. Quite a few of new notions have been introduced that justly deserve their separate drawers in the nomenclature chest, such as skyrmionic orders, for instance. What has been lacking, in our opinion, is a comprehensive classification, preferably based on symmetry criteria, but also on the observable manifestations, of different magnetic orders – after all, we scientists are Darwinian creatures as much as we are Linnaean, and we all know that naturally (dis)organized classification tend to grow into a Borges's categorization of animals[2]: 1. belonging to the Emperor, 2. embalmed ones, 3. trained ones, 4. suckling pigs, 5. mermaids, 6. fabulous ones, 7. stray dogs, 8. those included in these classification, 9. those that tremble insanely, 10. innumerable ones, 11. painted with a camel hair brush, 12. others, 13. those that have just broken a vase, and 14. those look like flies from afar. Category 6 is the one, seemingly, most popular with physics authors!

With this inspiration, when discussing the dilemma of introducing the new class of altermagnets, or demanding that the authors present them as a subset of standard antiferromagnets (as some referees suggested), we dared to draw our own classification of animals, as presented in Fig. 1. Some clarifications are in place. In this tree diagram, first, we divide all materials into magnetic and nonmagnetic. As trivial as this distinction is, it also requires a rigorous definition, and that is: magnets break time-reversal Kramers symmetry at least locally, that is to say, at least at some points within the material the local magnetic moment is non-zero. The next bifurcation separates collinear magnets, where the direction axis (but not necessarily the sign) of the local magnetization vector is the same through the entire crystal, from noncollinear. The latter can, in the same spirit, be classified as coplanar, where magnetization rotates in the real space, but remains in the same plane, and noncoplanar. Further subdivision could be meaningfully pursued (zero vs. nonzero vector or scalar chirality, toroidicity, etc; see a nice review by Cheong[3]).

The other branch of this taxonomy, which we shall dwell on a bit longer, is collinear magnets, of which we can split off, again by symmetry, incommensurate, commonly known as spin density waves, and commensurate magnetic structures. The latter can be compensated by symmetry, or not. The difference is fundamental in terms of the symmetries broken, but, importantly, there is a class of materials that do have a strictly zero net magnetic moment, but the compensation is not by symmetry, as there is no symmetry operation transforming one magnetic sublattice into the other. The physical meaning of magnetic compensation in this case can be understood by looking at a ferromagnetic semiconductor: by virtue of the Luttinger theorem, it can only have an integer magnetic moment per unit cell: 0, 1, 2 $\mu_B$, etc. That is to say, if it is zero, it is strictly zero, not just some small number, and this zero is robust: a small perturbation, such as pressure or stress, is not going to do away with this zero. Such "Luttinger-compensated" ferrimagnets is an interesting and largely overlooked class in its own.

But, for the purpose of these notes, we turn to the other branch, materials that are magnetically compensated by symmetry: there is a crystallographic symmetry group operation mapping one spin sublattice onto the other. Classical "Neel" antiferromagnets belong to this class, and till a few years ago it was believed that they are all more or less the same and distinctly different from ferro- and ferrimagnets (whether compensated or not). The following Perspective explains that a next stage of symmetry distinctions can and ought to be introduced, depending on whether this "mapping" operation preserves the symmetry in the momentum space (such operations are translations and inversions), or not (all others). Such additional classification is fairly fundamental: it is as important in the momentum space as the antiferro – ferro dichotomy is in the coordinate space. Not surprisingly, many observables are as sensitive to the symmetry breaking in the momentum space as to the same symmetry breaking in the reciprocal space, and, as explained in the Perspective, such "altermagnets" share more common properties with ferromagnets than with antiferromagnet (see Table 1). One can argue that they should be classified, if anything, as a subset of ferrimagnets rather than of antiferromagnets, but that, of course, defies any normal logic because symmetry-determined exact magnetic compensation clearly makes it impossible to call them ferromagnets.

Thus, we have concluded that in this particular case generating a new entity is not a whim, but a necessity – in reverence to John Punch's philosophy.

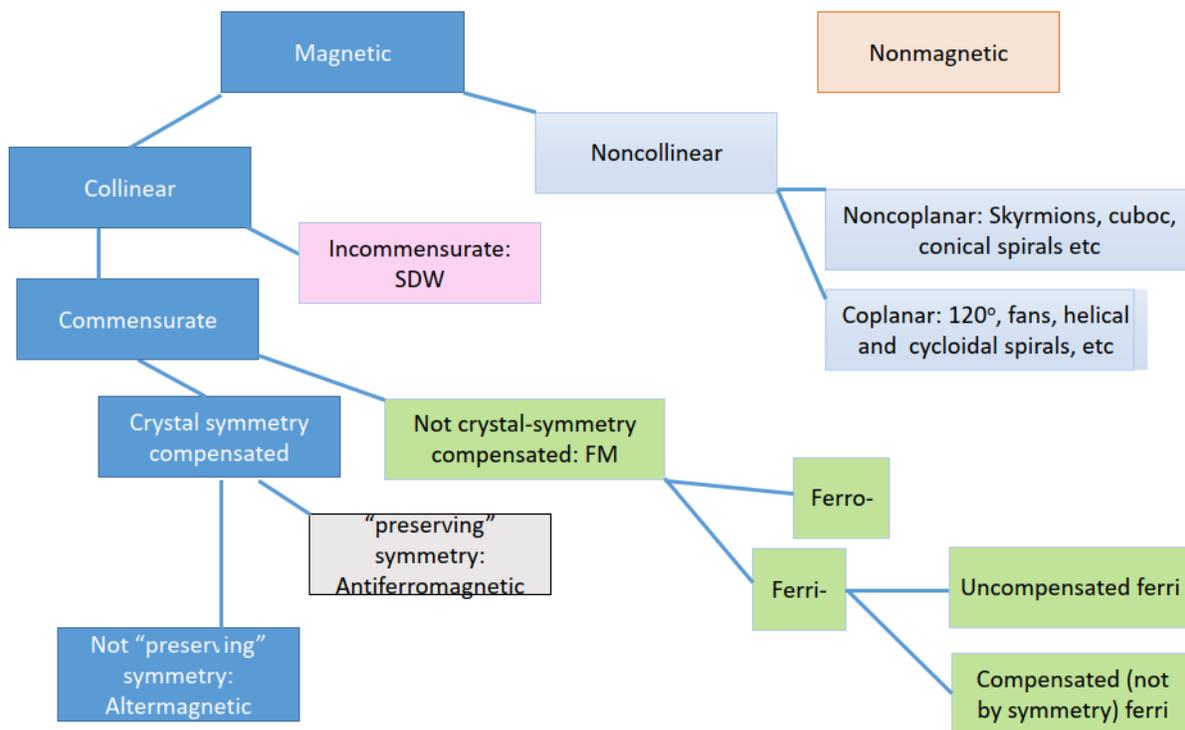

|  | FM | AF | AM |
|---|---|---|---|
| Net magnetization | Nonzero or zero | Zero | Zero |
| Kramers degeneracy | No | Yes | No |
| Anomalous Hall | Yes | No | Yes |
| Magnetooptics | Yes | No | Yes |
| Spin-current splitting | Yes | No | Yes |
| Suppression of Andreev reflection: diffuse contact | Yes | No | No |
| Suppression of Andreev reflection: ballistic contact | Yes | No | Yes |
| Supports singlet superconductivity | No | Yes | No |
| Supports locally (in k-space) unitary triplet superconductivity | No | Yes | No |
| Supports k-averaged unitary triplet superconductivity | No | Yes | Yes |
| Tunneling magnetoresistance | Yes | No | Yes |